# INVERSE REINFORCEMENT LEARNING CONDITIONED ON BRAIN SCAN

*Tofara Moyo*
*Mazusa*,C.E.O.
Bulawayo, Zimbabwe
tofaramoyo@gmail.com

*Abstract—* I outline a way for an agent to learn the dispositions of a particular individual through inverse reinforcement learning where the state space at time t includes an fMRI scan of the individual, to represent his brain state at that time. The fundamental assumption being that the information shown on an fMRI scan of an individual is conditioned on his thoughts and thought processes. The system models both long and short term memory as well any internal dynamics we may not be aware of that are in the human brain. The human expert will put on a suit for a set duration with sensors whose information will be used to train a policy network, while a generative model will be trained to produce the next fMRI scan image conditioned on the present one and the state of the environment. During operation the humanoid robots actions will be conditioned on this evolving fMRI and the environment it is in.

## INVERSE REINFORCEMENT LEARNING

Inverse reinforcement learning considers the process of recovering a reward function from a policy. The IRL problem statement can be framed as follows. [1].

Given 1) measurements of an agent's behavior over time, in a variety of circumstances, 2) measurements of the sensory inputs to that agent; 3) a model of the physical environment (including the agent's body).
Determine the reward function that the agent is optimizing.

This assumes that the demonstrator's actions are based on a Markov Decision Process. Markov decision processes may be visualized as a tuple (X, U, P, D, R). Where X is a finite set of states (here we consider the state as the environment that the agent is in) , U is a set of control inputs, P is a set of state transitions probabilities; D is the initial-state distribution from which the initial state x0 was drawn; and " R : X => R" is the reward function. The goal of IRL is to recover the unknown reward function R(X) .A good example of inverse reinforcement learning would be self-driving cars. Here the demonstrator is a human driver while the system tries to recover the reward function from the human's responses to external stimuli while driving. The MDP has X= the state of the environment surrounding the car at any given time. U is the responses given by the human driver to certain states of the environment. The goal of the system is to learn to respond in a similar manner to external stimuli. More sophisticated approaches have also been incorporated into this basic model, but the above description shows the elements that are similar to the system we wish to describe. Although within many IRL systems, a linear reward function may describe the agent's behavior sufficiently, a non-linear one may be optimal [2]. For complex tasks such as knot tying. If the reward function is assumed to be linear towards the feature the reward function becomes:

$$R(x) = w^T \phi(x) \tag{1}$$

Where R(x) is the reward for state x, w is a weight vector and $\phi(x)$ is the feature vector of state x.

Deep IRL approaches have been demonstrated to be useful in learning complex non-linear cost functions. [3]. Using back propagation or one of its variants they learn to map a state of the environment x to a set of control signals y. In this paper we will be using two deep architectures to illustrate the system. A generative model whose purpose will be to generate the brain fMRI sequence that is conditioned on $x_{t-1}$, and a deep feed forward network to map the output of the generative model at time *t* and the state $x_t$ to control $u_{t+1}$. We use a simple feed forward architecture here because the features of the output of the generative model have semantic meaning relative to the whole image rather than representing the same feature regardless of transformation. A simple fully connected feed forward deep network will do well to represent the information thus found within the image given the nature of information shown on an fMRI scan.

## GENERATIVE MODELS

Generative models aim to recover the probability distribution from where a data distribution is from with the purpose of generating novel instances of the data from that distribution.. They can be broken into two classes; explicit density estimators and implicit density estimators. A famous example of implicit density estimators are GANs, Generative adversarial networks. We shall concern ourselves though, for the purposes of simplicity, to an implicit density estimator in the form of Pixel LSTM RNN's. Other choices could have been pixel CNN's .Pixel RNN's show however to be more accurate in operation though they may be slower to train. A pixel RNN learns the joint distribution of a sequence of pixels. (LSTM pixel RNN's can learn long term dependencies and are the model of choice)

$$p(\mathrm{x}) = \prod_{i=1}^{n^2} p(x_i | x_1, ..., x_{i-1}) \qquad (1)$$

The value $p(x_i | x_1, ..., x_{i-1})$ is the probability of the *i*-th pixel $x_i$ given all the previous pixels $x_1, ..., x_{i-1}$. The generation proceeds row by row and pixel by pixel. Each pixel xi is in turn jointly determined by three values, one for each of the color channels Red, Green and Blue (RGB). The distribution

$$p(x_i/\mathrm{x}_{<i}) = p(x_i/x_{<i})p(x_i,G/x_{<i},x_i,R)p(x_i,B/x_{<i},x_i,R,x_i,G) \qquad (2)$$

[4].The generation of an image is sequential and begins on a predetermined point on the image (usually
one of the corners) depending on the manner the network was trained. One distinction is between diagonal biLSTM's and row LSTM's. Pixel RNN's may also be conditional. This means that their output may depend on a particular label or embedding. They are often used to complete images that contain occlusions. This is very important for the development of the system at hand.We will need a Pixel RNN to produce the next image in a sequence of fMRI scans conditioned on both the previous fMRI scan and the current state x of the environment.

$$F_{t+1}|F_t, x_t \qquad (3)$$

## FMRI SCANS

An fMRI is the development of the MRI scan that shows many advantages over it for the purposes of this system. While a magnetic resonance image MRI shows the contrast in densities between different forms of matter within the brain, the fMRI or functional MRI displays in the form of a video the metabolic processes within the brain. [5].It does this by monitoring the prevalence of oxygen within the blood vessels of the brain. This is important for us because an important assumption of this paper is that the metabolic processes of the brain are conditioned on the thoughts and thought processes of the individual taking the scan. We will use this information to create a system that will condition its behavior on a generated version of an fMRI slide a time t-1 and the environment at time. Why would we expect that this would be a worthwhile endeavor? Well the assumption is that the world functions in a deterministic manner. At any one time we only have access to the information present at that time. This is the Markov property of the Universe, and that state of the universe fully determines the next. That implies that memory and thought processes of an individual are also fully deterministic, conditioned on the rest of the universe. Here by Universe, we are really concerned with the state *x* at time *t*. Because of the Markov property, in order to ever know what the individual taking the scan would do in any situation we just need to know $(F_t, x_t)$. That is, his brain state at that time and the state of the environment he is in. This includes all behavior no matter the complexity, even if it is heavily dependent on long term memory, as the information needed from it to form the next control signal will be present within the brain state at time *t*.

## THE FINAL EXPOSITION

The final system explained takes the form of a model with the following elements. Neural networks to gather sensory information. A microphone for the purposes of recoding the human agent's voice and a neural network whose final purpose it is to output a vocal signal as the output to a control signal. A portable fMRI scan. (This may be a low resolution model whose output would be up resolved before the back propagation procedure) A deep fully connected feed forward network connected to the output of the fMRI scan at time t and the output of the neural networks conditioned on sensory information as input, and as its output, the control signals to the neural network controlling the speaker representing a voice and the motors of the robot. (This is a system where the motion sensors will approximate where they should go given the actual movement of the human agent conditioned on the fact that they may have fewer degrees of freedom.) The elements will take on their roles within a "suit" to be worn by the individual and all the networks will learn using the backpropagation procedure. The fMRI network will produce the material that will be used to train a pixel RNN to generate time series data where the next brain state is produced from the past one conditioned on the present state of the environment. We will typically only collect data through the duration that the suit is worn, then perform the training of the networks at another stage. During operation we should expect the humanoid robot to function as if it were the individual who it was trained on as the fMRI sequence produced in it would presumably follow the dispositions of the individual.

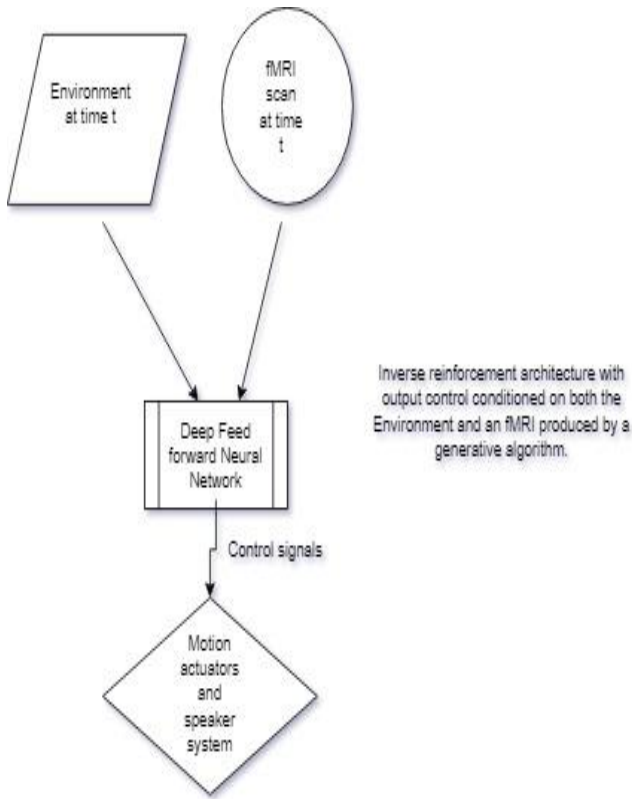

Inverse reinforcement architecture with output control conditioned on both the Environment and an fMRI produced by a generative algorithm.

*Fig1. Basic Architecture*

## CONCLUSION

It can be seen that this system can model long-term memory without the need of recurrence. It could be used as an adjunct to self-driving cars, where the fMRI of each expert driver's brain is taken to condition the control signals U on. As of yet it may still not be feasible to perform experiments on this approach. A notable drawback is the absence of the invention of portable fMRI scanners. Also the generative process may not be developed in the literature to the stage where it can model the long-term dependencies in the fMRI slides faithfully. And the generation process itself described here may be too slow to put in a Robot expected to perform well on runtime. Nevertheless this promises to be an interesting avenue to follow.

ACKNOWLEDGMENT

I would like to acknowledge the work of all the people cited as well as the lectures who taught me during my time as an undergraduate at the National University of Science and Technology N.U.S.T. in Zimbabwe.

REFERENCES

[1] Russell, Andrew Y. Ng, Algorithms for inverse reinforcement learning [2000].
[2] Levine et al. Non linear Inverse Reinforcement Learning with Gausssian processes [2011].
[3] Grubb and Bagnell , Bradley ,Boosted backpropagation learning for training deep modular networks [2010].
[4] Aaron van den Oord, Nal Kalchbrenner, Koray Kavukcuoglu , pixel recurrent neural networks [2016].
.
[5] Bandettini PA,Wong EC, Hinks RS, Tikofsky RS, Hyde JS Magn Reson Med. 1992 Jun; 25(2):390-7.